\documentclass[12pt,a4paper]{article}
\usepackage{amsmath}
\usepackage{amsfonts}
\usepackage{amssymb}
\usepackage{textcomp}
\usepackage[dvips]{graphicx}
\usepackage{epsfig}
\usepackage{bm}
\usepackage{dcolumn}
\usepackage{amsfonts}
\usepackage{color}
\usepackage[left=2cm,right=2cm,top=2cm,bottom=2cm]{geometry}
\usepackage[left=2cm,right=2cm,top=2cm,bottom=2cm]{geometry}
\usepackage[tableposition=top]{caption}
\usepackage{subcaption}
\DeclareCaptionLabelFormat{gostfigure}{Fig. #2}
\DeclareCaptionLabelSeparator{gost}{.~~}
\begin{document}
\title{\textbf{On Kasner solution in Bianchi I $f(T)$ cosmology}}
\author{Maria A. Skugoreva$^{1}$\footnote{masha-sk@mail.ru}, Alexey V. Toporensky$^{1,2}$\footnote{atopor@rambler.ru}\vspace*{3mm} \\
\small $^{1}$Kazan Federal University, Kremlevskaya 18, Kazan, 420008, Russia\\
\small $^{2}$Sternberg Astronomical Institute, Lomonosov Moscow State University,\\
\small Moscow, 119991, Russia}

\date{ \ }
\maketitle

\begin{abstract}
Recently cosmological dynamics of anisotropic Universe in $f(T)$ gravity became an area of intense investigations. Some earlier papers devoted to this issue contain contradictory claims about the nature and propertied of vacuum solutions in this theory. The goal of the present paper is to clarify this situation. We compare properties of $f(T)$ and $f(R)$ vacuum solutions and outline differences between them. The Kasner solution appears to be an exact solution for the $T=0$ branch, and an asymptotic solution for $T \ne 0$ branch. It is shown that Kasner solution is the past attractor if $T<0$, being past and future attractor for $T>0$ branch.
\end{abstract}

Kasner solution, being one of the first known exact solution in relativistic cosmology \cite{Kasner} continues to be one of the most important exact solution in General Relativity (GR) or its modifications. One of the reasons is that despite this is a vacuum solution, it is a good approximation near a cosmological singularity for almost all matter sources (except for a stiff fluid) in a flat anisotropic Universe. Moreover, a general cosmological singularity is believed to be constructed as an infinite series of consecutive epochs each of them being a particular Kasner solution with a good accuracy (though a mathematical prove of this scenario is still absent in full details -- see, for example, \cite{Uggla}) --- the famous Belinskii-Khalatnikov-Lifshitz (BKL) scenario \cite{BKL}. So that, Kasner set of solutions provides "building blocks" for BKL picture.

If we assume that GR needs some modifications at UV scale, it is natural to expect that such modifications should change the behaviour near a cosmological singularity significantly. That is why the fate of Kasner solution in modified gravity theories is an area of intense investigations. A lot of efforts have been devoted to Kasner solutions and its modifications in quadratic gravity. We remind a reader that Kasner solution is a solution for an anisotropically expanding Universe with scale factors changing as powers of time. These power exponents are subject of two conditions giving us their sum as well as the sum of their squares (both sums are equal to unity). In quadratic gravity, two different situations were identified:
\begin{itemize}
\item If the equations of motion are of the second order, as in GR (that is, in Gauss-Bonnet gravity), the power-law solution for scale factor is an asymptotic solution. In the high-curvature regime, these two conditions for power exponents are different from those in GR Kasner solution \cite{Deruelle1,Deruelle2,TT}, while GR Kasner solution being an asymptotic solution in the low-curvature regime.
\item In fourth order gravity (like $R+R^2$ or a general quadratic gravity) the Kasner solution (with the same conditions for the  exponents) is an {\it exact} vacuum solution. However, since the phase space has two additional dimensions in comparison with GR, Kasner solution in quadratic gravity may be in some situations unstable \cite{Hervik, Muller}.
\end{itemize}

Recently new class of modified gravity theories has started to attract great attention. It is based on the Teleparallel Equivalent to General Relativity (TEGR) --- a theory first considered by Einstein in 1920-th \cite{Unzicker:2005in} where the Levi-Civita connection (torsion free, non-zero curvature) have been replaced by Weitzenb\"{o}ck connection \cite{Weitzenbock} (curvature-free, non-zero torsion), and curvature scalar $R$ in the action by torsion scalar $T$. It appears that despite of different mathematical background, TEGR and GR are identical at the level of equations of motion. For a review see, for example, the book \cite{JGPereira}. Now it is known that in the cases more complicated than the standard Einstein-Hilbert action, the theory based on torsion has equations of motion different from the theory based on curvature. In particular, $f(T)$ theory is not equivalent to $f(R)$ theory, if the function $f$ is not a linear function. 

This motivated studies of cosmological dynamics in $f(T)$ gravity. Recently many papers on this topic have been appeared mostly concentrating on FRW cosmology (see, for example, \cite{Nunes1, Nunes2, Jarv, Saridakis} and references therein. Anisotropic cosmology is the next natural step in this direction. However, $f(T)$ theory has its own problems connecting with the lack of local Lorentz invariance \cite{Lorentz1, Lorentz2}. This lead to a situation when not all tetrads corresponding to the chosen metric form gives us the correct equations of motion \cite{tet1,tet2}, so a separate problem to choose a so called ``proper tetrad'' \cite{Ferraro} appears. Alternatively, a proper non-zero spin connections must be associated with a given tetrad \cite{Krssak}. Careful investigation of this problem in connection with anisotropic metrics is still missing. In such a situation we can use an heuristic argument of \cite{Krssak} that if a tetrad (or spin connection) is chosen in a bad way, resulting equation of motion should be (in some sense) pathological. The most common pathology is a requirement that the second derivative of $f$ with respect to $T$ vanishes, which evidently brings us back to TEGR. So, a reasonable current strategy may be in choosing the simplest tetrad associated with Bianchi I metric, and study corresponding equations of motion, if they do not show such pathologies.

    This strategy have been implemented recently in several papers \cite{Rodrigues,Cai,Barrow1,Barrow2}. The resulting equations of motion appear to be nonpathological. These papers have, however, some contradictory statements. In particular, the study \cite{Rodrigues} concludes, that a vacuum solution exists only for a particular $f(T)=\sqrt{-T}$ theory and it must be isotropic. On the contrary, the paper \cite{Barrow2} claims that Kasner solution is still solution for $f(T)$ theory, though it is unstable. The fact that Kasner solution remains to be a solution can be easily checked by direct substitution to the equations of motion. So that, it seems that the second above mentioned alternative could realize. However, the problem is that once the tetrad is fixed, the number of degrees of freedom in $f(T)$ theory and in TEGR is the same. In $f(R)$ gravity the instability of Kasner solution (which is the exact solution in that theory) is due to extra degrees of freedom, which are absent in the $f(T)$ theory in question. This contradiction needs a careful analysis, which is the goal of the present paper. We will see below that none of two quadratic gravity alternatives regarding Kasner solution can be true for $f(T)$ cosmology where we meet a third, different situation.

    We consider cosmological models in modified teleparallel gravity $f(T)$, where (as in TEGR) the dynamical variables are tetrad fields ${\mathbf{e}_A(x^\mu)}$; here Greek indices are space-time and capital Latin indices relate to the tangent space-time. The metric tensor is given by $g_{\mu\nu}=\eta_{\mathrm{AB}}\, e^A_\mu \, e^B_\nu,$
where $\eta_{\mathrm{AB}}=\mathrm {diag} (1,-1,-1,-1)$.  

    The action of $f(T)$ theory without matter has the form
\begin{equation}
S=\frac{1}{16\pi G}\int e~ \left( f(T)\right)d^{4}x,
\label{action}
\end{equation}
where $e=\sqrt{-g}$ is the determinant of the tetrad, ~~$f(T)$ --- a general differentiable function the torsion scalar $T$. Units $\hbar=c=1$ will be used.  

    We choose the following diagonal tetrad 
\begin{equation}
\label{tetrad}
e^A_\mu = \mathrm{diag}(1, a(t), b(t), c(t)),
\end{equation}    
which relates to the Bianchi I metric ~~$\mathrm{d}s^2=\mathrm{d}t^2-a^2(t)\mathrm{d}x^2-b^2(t)\mathrm{d}y^2-c^2(t)\mathrm{d}z^2$,~~ where ~~$a(t)$, ~~$b(t)$,~~ $c(t)$~~ are scale factors. The torsion scalar for the chosen tetrad (\ref{tetrad}) is
\begin{equation}
\label{Tabc}
T=-\frac{2}{abc}(c\dot a \dot b+b\dot a \dot c +a\dot b \dot c),
\end{equation}
where a dot denotes the derivative with respect to time. We can rewrite the expression for $T$ (\ref{Tabc}) using anisotropic Hubble parameters $H_a\equiv\frac{\dot a}{a}$, ~~$H_b\equiv\frac{\dot b}{b}$, ~~$H_c\equiv\frac{\dot c}{c}$
\begin{equation}
\label{THabc}
T=-2(H_a H_b+H_a H_c+H_b H_c),
\end{equation}
which reduces to $T=-6H^2$ in the isotropic case $a(t)=b(t)=c(t)$, ~~$H_a=H_b=H_c=H$.

Varying the action (\ref{action}) with respect to the chosen tetrad (\ref{tetrad}) equations of motion are obtained (see, for example \cite{Rodrigues})
\begin{equation}
\label{constraint}
-2 T f_T+f(T)=0,
\end{equation}
\begin{equation}
\label{system1}
\dot T f_{TT}(H_b+H_c)+\frac{f}{2}+f_T\left( \dot H_b+\dot H_c +{(H_b)}^2+{(H_c)}^2+2 H_b H_c+H_a H_b+H_a H_c\right)=0, 
\end{equation}
\begin{equation}
\label{system2}
\dot T f_{TT}(H_a+H_c)+\frac{f}{2}+f_T\left( \dot H_a+\dot H_c +{(H_a)}^2+{(H_c)}^2+2 H_a H_c+H_a H_b+H_b H_c\right)=0,
\end{equation}
\begin{equation}
\label{system3}
\dot T f_{TT}(H_a+H_b)+\frac{f}{2}+f_T\left( \dot H_a+\dot H_b +{(H_a)}^2+{(H_b)}^2+2 H_a H_b+H_a H_c+H_b H_c\right)=0.
\end{equation}
Here we denote $f_T=\frac{d f(T)}{d T}$, ~~ $f_{TT}=\frac{d^2 f(T)}{dT^2}$. 

    We investigate cosmological models with the Lagrangian density function $f(T)=T+f_0 T^N$, where $f_0$, ~~$N$ are parameters. It is important to notice that the constraint equation (\ref{constraint}) in this model is an algebraic relation for the torsion scalar $T$. This means that, in contrast to $f(R)$ theory where there exists a differential equation for the curvature scalar (see, for example, \cite{Angelo}), and, so, $R$ has a dynamics, the torsion scalar at the entire trajectory can be equal to only one value of a discrete set of possible values. In particular for the power-law model studied in the present paper $T$ can belong to only two branches of solutions. Substituting $f(T)=T+f_0T^N$ and $f_T(T)=1+Nf_0T^{N-1}$ to (\ref{constraint}) we find
\begin{equation}
T\left( f_0(1-2 N)T^{N-1}-1\right)=0,
\end{equation}
then
\begin{equation}
\begin{array}{l}
\label{Tconst}
\textbf{1).}~~ T=0.\\
\textbf{2).}~~ T^{N-1}=\frac{1}{f_0(1-2 N)}=const, ~~N\neq\frac{1}{2}.
\end{array}
\end{equation}    
\\
\\\textbf{1).~~ The case of} ~~$T=0$.
\\

~~~If ~~$T=0$~~ then ~~$f_T=1$~~ and the field equations (\ref{system1})-(\ref{system3}) have the form
\begin{equation}
\label{system11}
\dot H_b+\dot H_c +{(H_b)}^2+{(H_c)}^2+2 H_b H_c+H_a H_b+H_a H_c=0, 
\end{equation}
\begin{equation}
\label{system21}
\dot H_a+\dot H_c +{(H_a)}^2+{(H_c)}^2+2 H_a H_c+H_a H_b+H_b H_c=0,
\end{equation}
\begin{equation}
\label{system31}
\dot H_a+\dot H_b +{(H_a)}^2+{(H_b)}^2+2 H_a H_b+H_a H_c+H_b H_c=0.
\end{equation}
This system of field equations coincides with that in General Relativity (GR) for the Bianchi I metric. Therefore, the Kasner solution \cite{Kasner} ~~$a(t)=a_0 t^{p_1}$, ~~$b(t)=b_0 t^{p_2}$, ~~$c(t)=c_0 t^{p_3}$~~ with ~~$p_1+p_2+p_3=1$,~~ ${(p_1)}^2+{(p_2)}^2+{(p_2)}^2=1$ is the exact solution for this system. Moreover, as in GR it is a general vacuum solution for the Bianchi I metric.

    It is easy to see that there is no de Sitter solution with $H_a=H_b=H_c=H_{dS}=const\neq0$ on this branch.
    
    Since the constraint equation is an algebraic, but not a differential equation for $T$ it is impossible to vary $T$, setting it to some non-zero value on the branch in question. So, all corrections to GR vanishes on this branch of Bianchi I vacuum solutions. This situation has no analogs in both $f(R)$ and Gauss-Bonnet gravity. Only if matter sources are taken into account, corrections to GR become non-vanishing (in this case $T$ evidently is not zero). That is why the question of stability of vacuum Kasner solution on $T=0$ branch is a meaningless question, because this branch contains no other vacuum solutions.
\\
\\
\\\textbf{2).~~ The case of} ~~$T^{N-1}=\frac{1}{f_0(1-2 N)}=const$.
\\

~~~For ~~$T^{N-1}=\frac{1}{f_0(1-2 N)}=const$~~ we get ~~$f_T=\frac{1-N}{1-2N}$ and from (\ref{constraint}) $f=2 T f_T=2 T\frac{1-N}{1-2N}$.~~ Then equations of motions (\ref{system1})-(\ref{system3}) reduce to  
\begin{equation}
\label{system12}
\frac{1-N}{1-2N}\left[T+\left( \dot H_b+\dot H_c +{(H_b)}^2+{(H_c)}^2+2 H_b H_c+H_a H_b+H_a H_c\right)\right]=0, 
\end{equation}
\begin{equation}
\label{system22}
\frac{1-N}{1-2N}\left[T+\left( \dot H_a+\dot H_c +{(H_a)}^2+{(H_c)}^2+2 H_a H_c+H_a H_b+H_b H_c\right)\right]=0,
\end{equation}
\begin{equation}
\label{system32}
\frac{1-N}{1-2N}\left[T+\left( \dot H_a+\dot H_b +{(H_a)}^2+{(H_b)}^2+2 H_a H_b+H_a H_c+H_b H_c\right)\right]=0.
\end{equation}

    In the limit $|H_a|\rightarrow \infty$, $|H_b|\rightarrow \infty$, $|H_c|\rightarrow \infty$, we can neglect the constant term $f/2$ and find the {\it asymptotic} Kasner solution ~~$a(t)=a_0 {|t-t_0|}^{p_1}$, ~~$b(t)=b_0 {|t-t_0|}^{p_2}$, ~~$c(t)=c_0 {|t-t_0|}^{p_3}$, $t\rightarrow t_0$~~ with ~~$p_1+p_2+p_3=1$,~~ ${(p_1)}^2+{(p_2)}^2+{(p_2)}^2=1$.
    
    Moreover, the exact de Sitter solution exists $H_{dS}=\pm\sqrt{-T/6}$ for the negative values of $T$. We write down conditions of an existence of de Sitter solution \cite{Barrow2}:
\\
\\\textbf{a).}~~ if $N$ --- even, then $f_0>0$,\\
\textbf{b).}~~ if $N$ --- odd, then $f_0<0$.    
\\    

\begin{figure}[hbtp]
\includegraphics[scale=0.44]{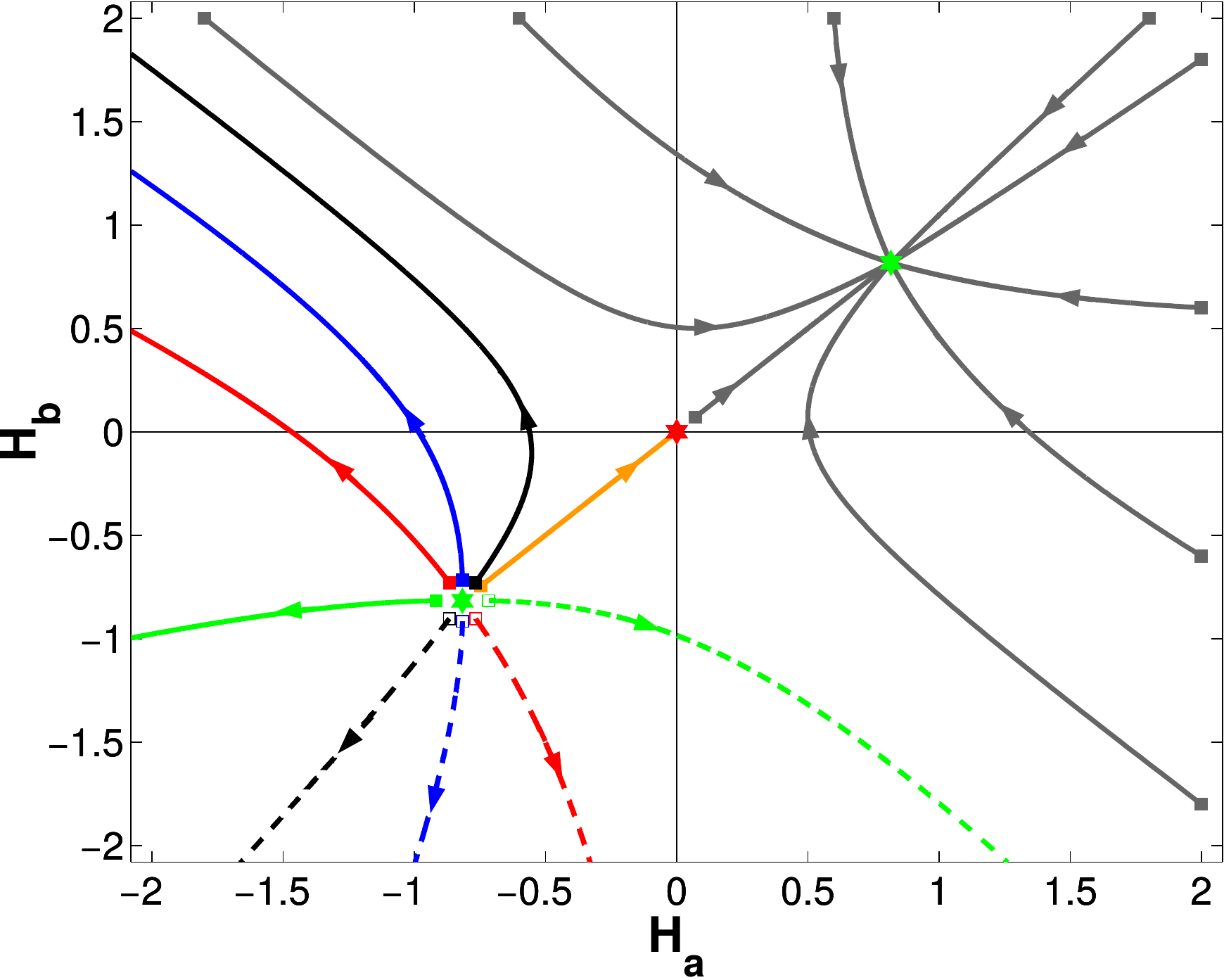}\qquad
\includegraphics[scale=0.44]{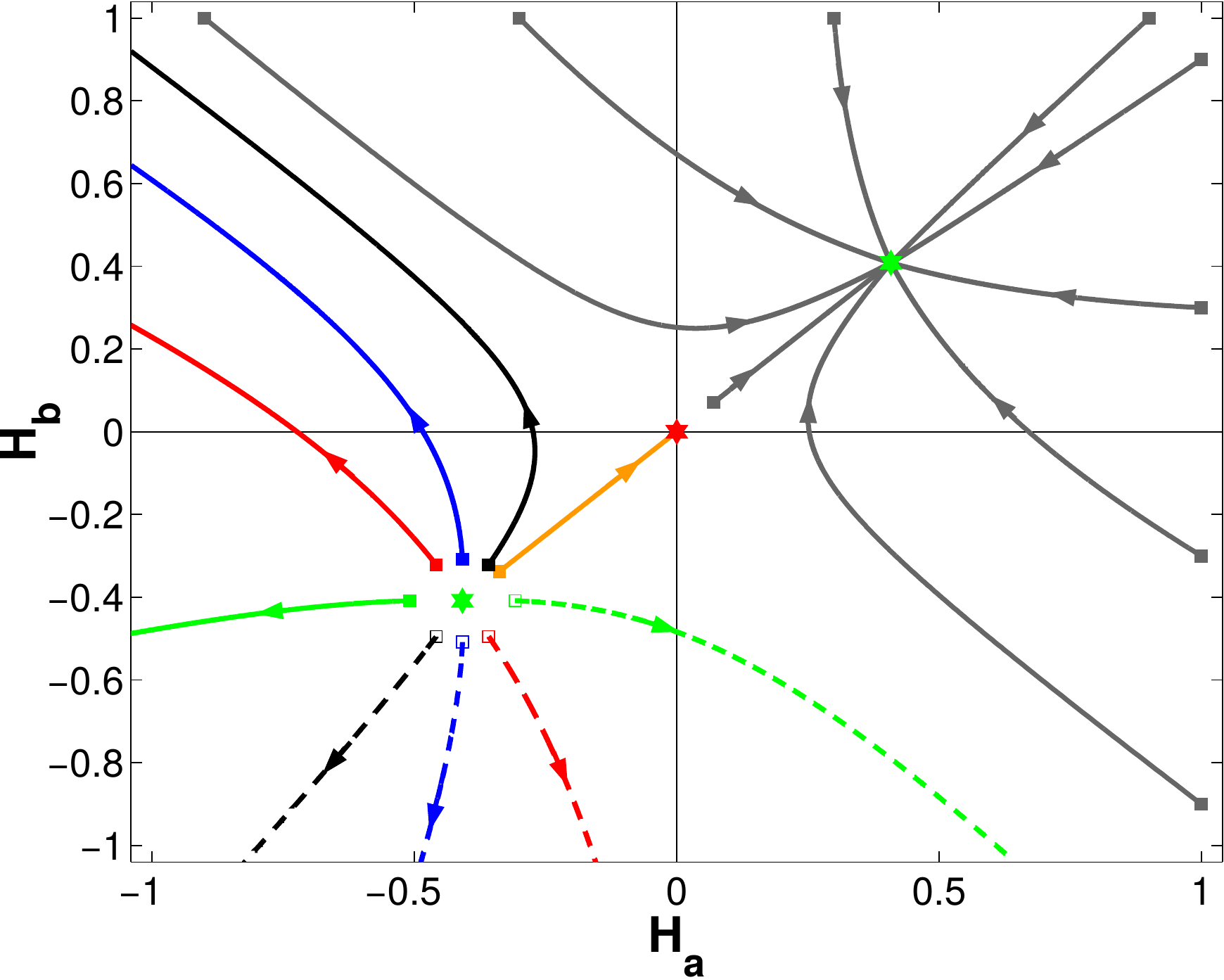} 
\caption{Phase portraits for $N=2$, $f_0=\frac{1}{12}$ (left) and $N=3$, $f_0=-\frac{1}{5}$ (right). Red and green stars denote the point ~$(0; 0)$~ and de Sitter ones ~$(H_{dS}; H_{dS})$~ correspondingly. Initial points $(H_a(0); H_b(0))$ have the following coordinates:
\\\textbf{1).}~~ for black, blue, red and green squares $H_a(0)$ equals to values from $H_{dS}-0.1$ to 
\\\text{~~~~~~~}$H_{dS}+0.1$, where $H_{dS}<0$, with step 0.05 and corresponding values 
\\\text{~~~~~~~}$H_b(0)=\pm\sqrt{{(0.1)}^2-{(H_a(0)-H_{dS})}^2}+H_{dS}$, 
\\\textbf{2).}~~ for the orange square $H_a(0)=H_b(0)=H_{dS}+0.1/\sqrt{2}$, where $H_{dS}<0$.
\\\textbf{3).} in the left plot for gray squares 
\\\text{~~~~~}\textbf{a).}$H_a(0)$ equals to values from $-1.8$ to $1.8$ with step $1.2$ and $H_b(0)=2$,
\\\text{~~~~~}\textbf{b).} $H_a(0)=2$ and $H_b(0)$ equals to values from $-1.8$ to $1.8$ with step $1.2$,
\\\textbf{4).} in the right graph for gray squares 
\\\text{~~~~~}\textbf{a).} $H_a(0)$ equals to values from $-0.9$ to $0.9$ with step $0.6$ and $H_b(0)=1$,
\\\text{~~~~~}\textbf{b).} $H_a(0)=1$ and $H_b(0)$ equals to values from $-0.9$ to $0.9$ with step $0.6$.}
\label{Fig1}
\end{figure}

     We apply the numerical integration of the initial system (\ref{system1})-(\ref{system3}) and check the constraint equation ~~$f_0(1-2 N)T^{N-1}-1=0$~~ at each step of the integration. It should be noted that we can use (\ref{Tconst}) to decrease the number of degrees of freedom. However, in GR numerics, especially on singular solutions, it is better to use the constraint equation for checking the numerical accuracy. This is a widespread practice. The numerical work is carried out for $N=2$ and $N=3$ with the initial values of Hubble parameters near the de Sitter point and we plot the evolution of trajectories going to the past, that is from the de Sitter universe to the Kasner one. This result is shown in Fig.~\ref{Fig1}. This figure appears to be the same as the corresponding figure of \cite{Barrow2}, so we disagree with this paper only in some points of interpretation, not in numerical results (see summary below in conclusions).    

\begin{figure}[hbtp]
\includegraphics[scale=0.44]{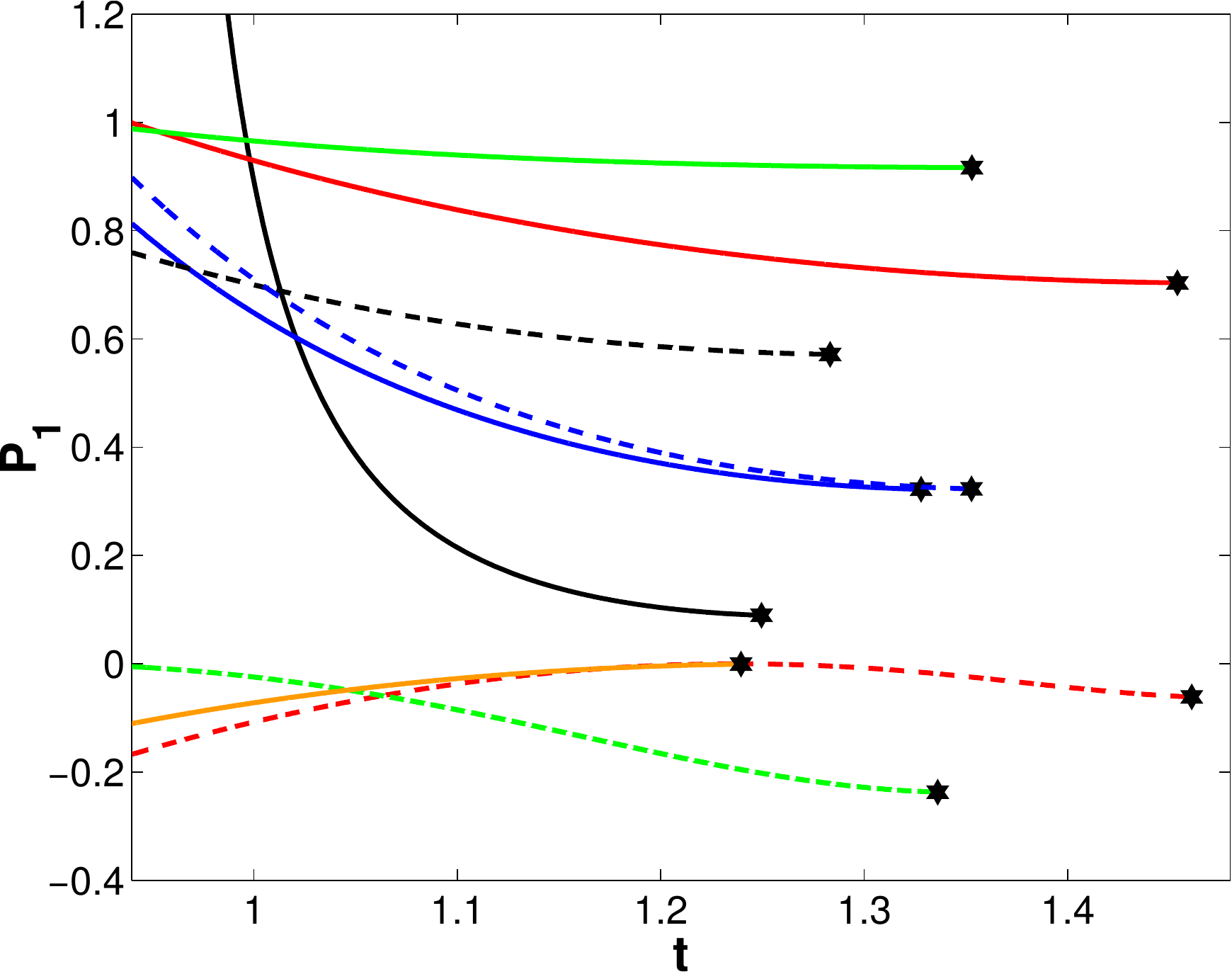}\qquad
\includegraphics[scale=0.44]{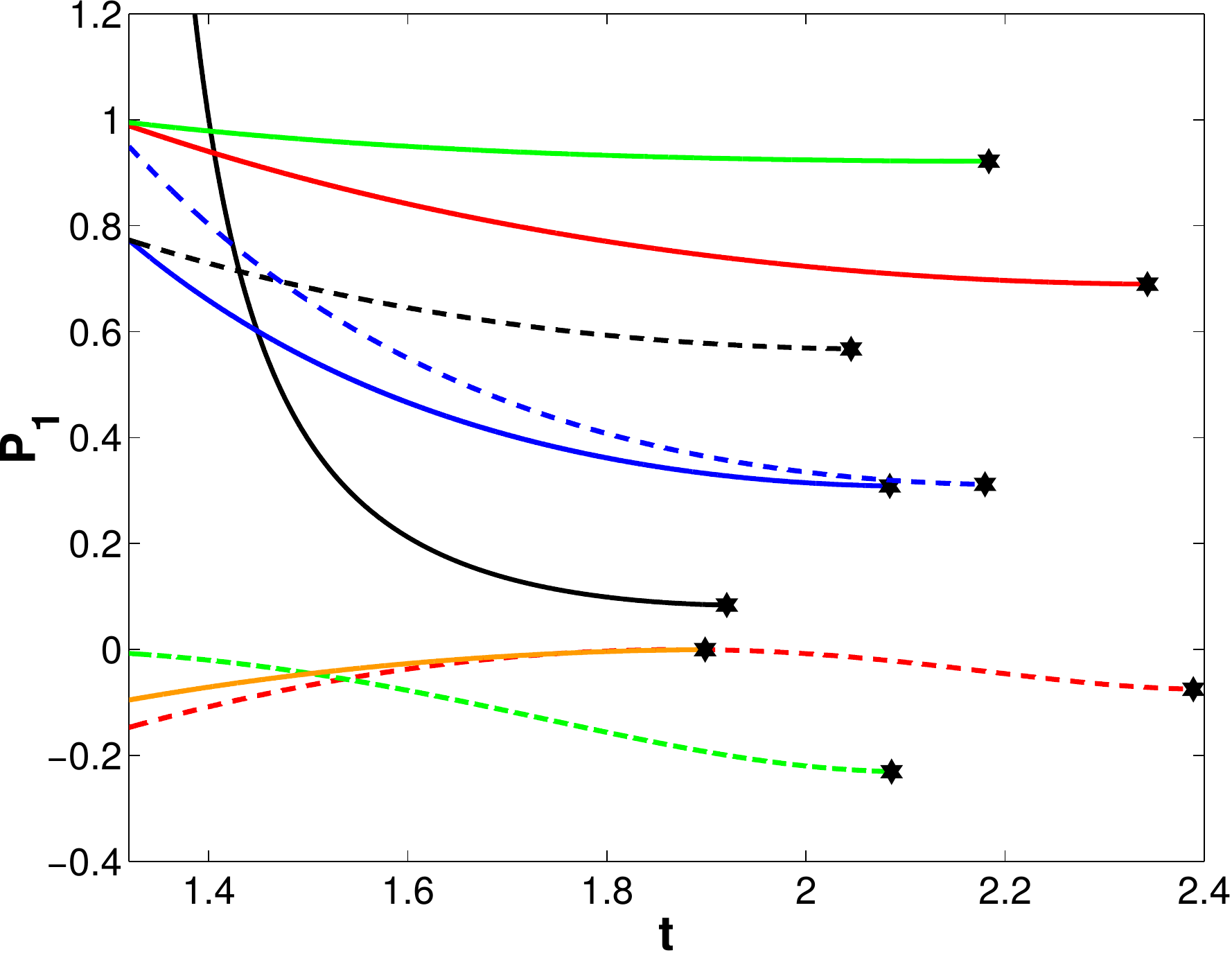} 
\caption{The final stage of the evolution of the dimensionless parameter $P_1(t)$ for $N=2$, $f_0=\frac{1}{12}$ (left) and $N=3$, $f_0=-\frac{1}{5}$ (right). Initial values of $H_a(0)$, $H_b(0)$ are chosen as in Fig. \ref{Fig1} (see the corresponding colors and  types of line). The black star denotes the moment of 
a cosmological singularity.}
\label{Fig2}
\end{figure}

\begin{figure}[hbtp]
\includegraphics[scale=0.44]{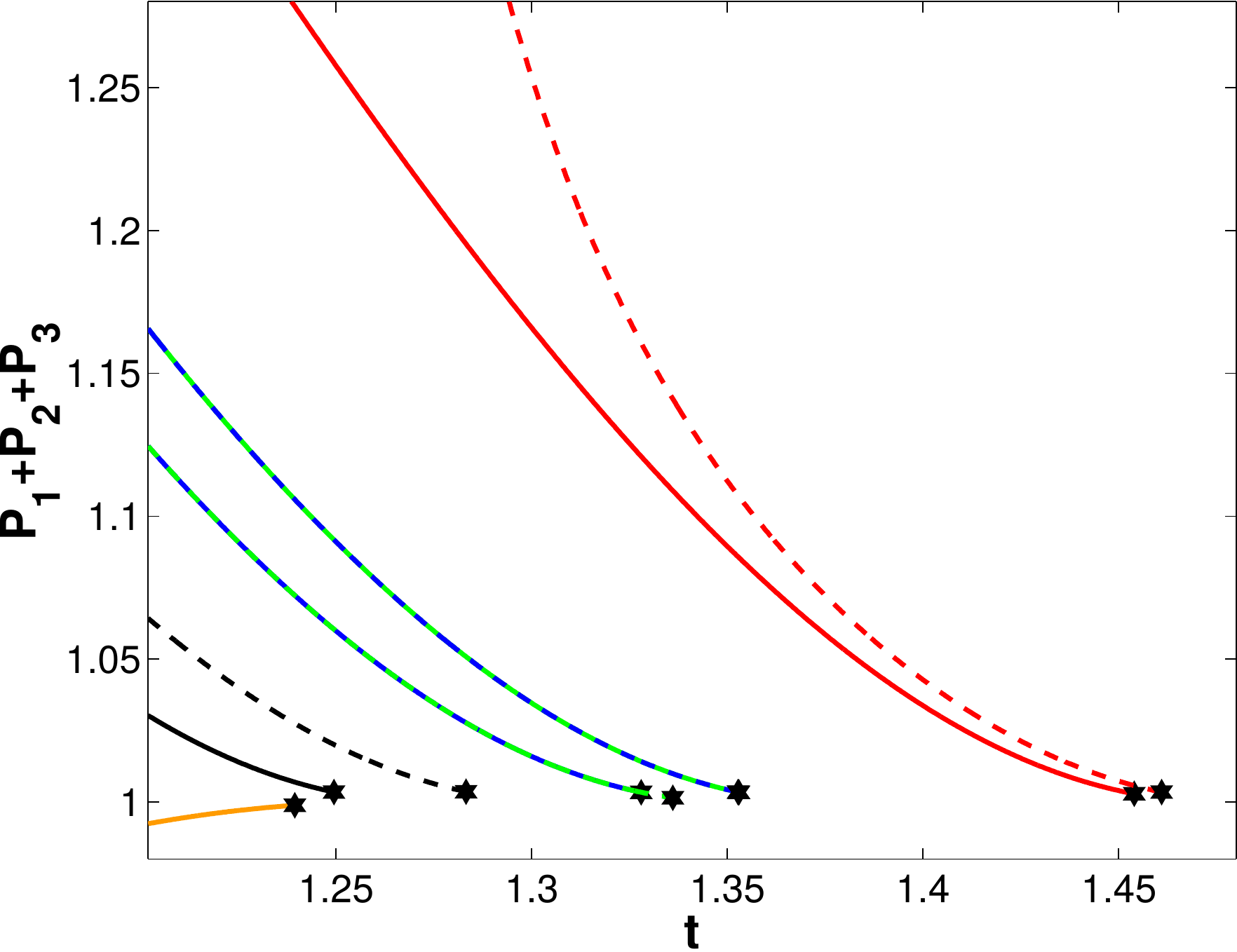}\qquad
\includegraphics[scale=0.44]{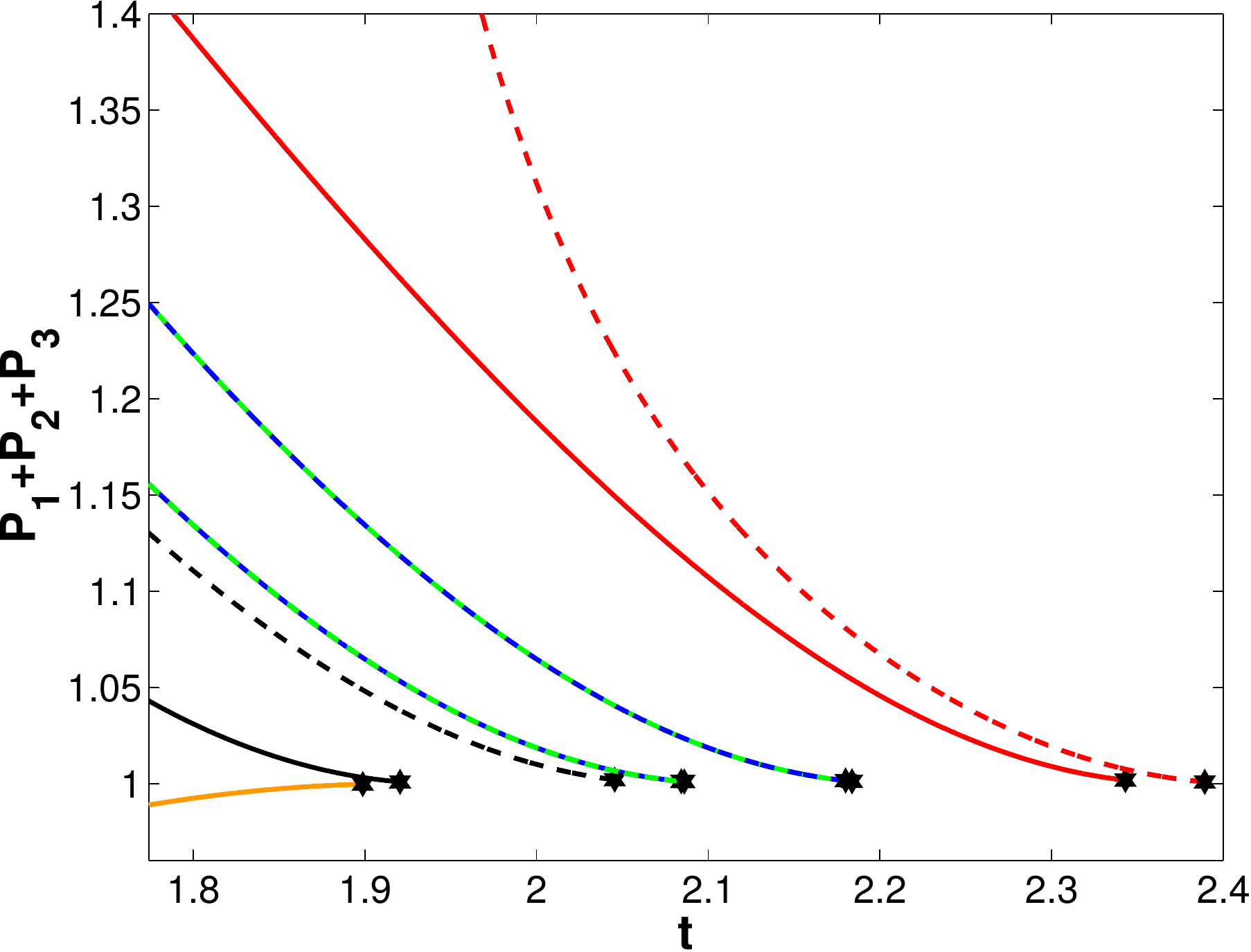} 
\caption{The final stage of the evolution of the sum $P_1+P_2+P_3$ for $N=2$, $f_0=\frac{1}{12}$ (left) and $N=3$, $f_0=-\frac{1}{5}$ (right). Initial values of $H_a(0)$, $H_b(0)$ are chosen as in Fig. \ref{Fig1} (see the corresponding colors and types of line). The black star denotes the moment of a cosmological singularity.}
\label{Fig3}
\end{figure}

\begin{figure}[hbtp]
\includegraphics[scale=0.44]{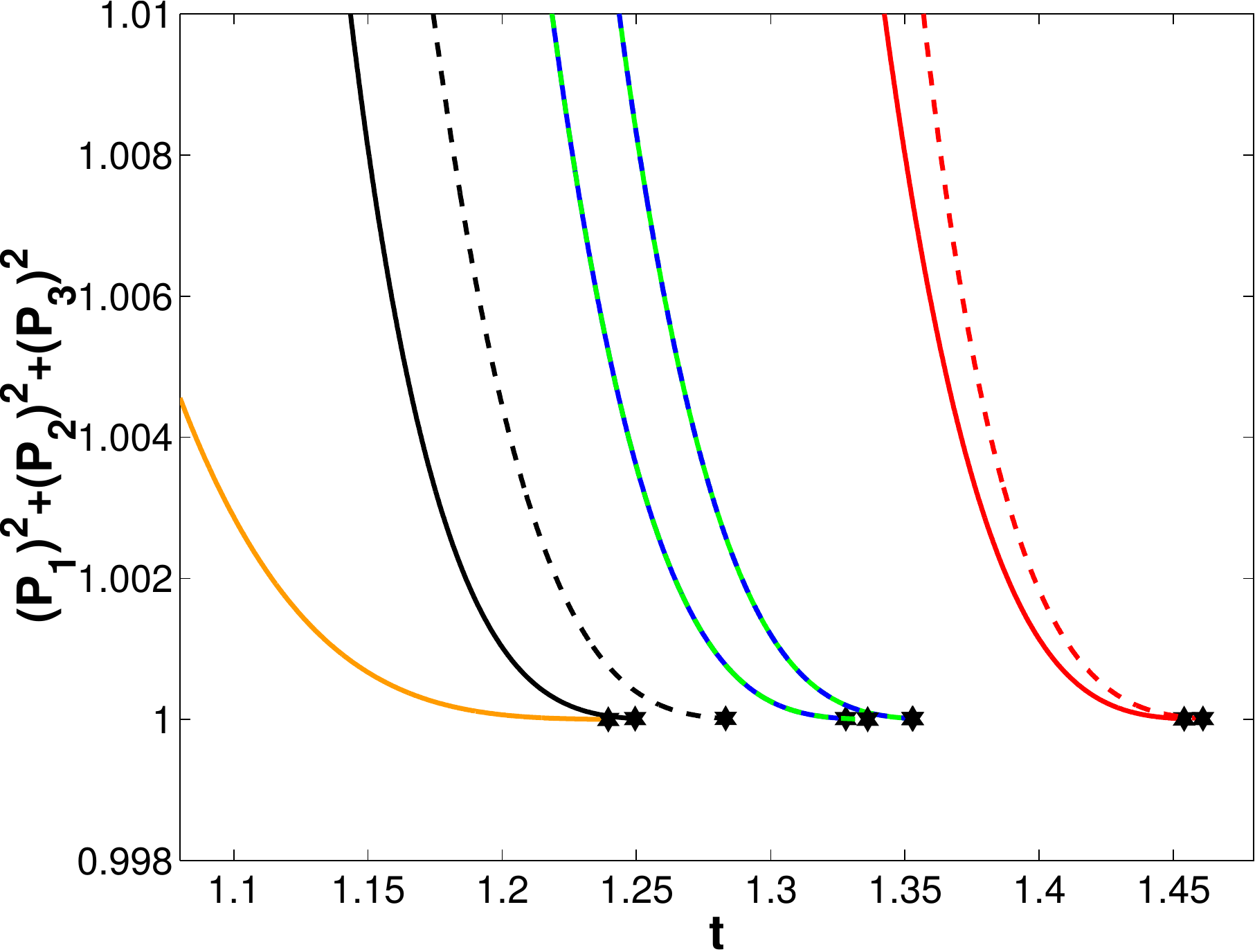}\qquad
\includegraphics[scale=0.44]{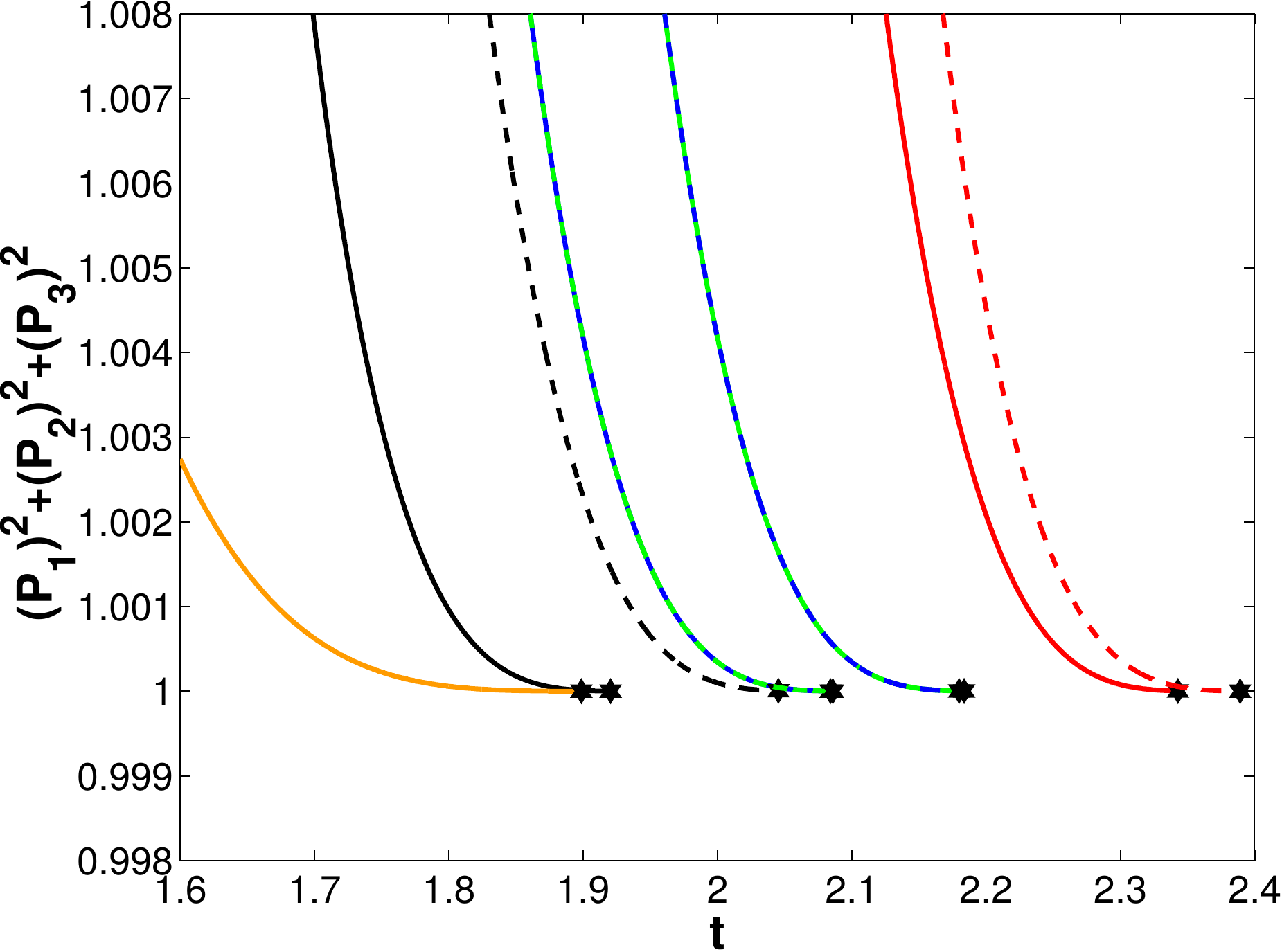} 
\caption{The final part of the evolution of the sum ${(P_1)}^2+{(P_2})^2+{(P_3)}^2$ for $N=2$, $f_0=\frac{1}{12}$ (left) and $N=3$, $f_0=-\frac{1}{5}$ (right). Initial values of $H_a(0)$, $H_b(0)$ are chosen as in Fig. \ref{Fig1} (see the corresponding colors and types of line). The black star denotes the moment of a cosmological singularity.}
\label{Fig4}
\end{figure}

    We also calculate the following dimensionless parameters
\begin{equation}
\label{p1p2p3}
P_1(t)=-\frac{{H_a}^2}{\dot {H_a}}, ~~~~P_2(t)=-\frac{{H_b}^2}{\dot {H_b}}, ~~~~P_3(t)=-\frac{{H_c}^2}{\dot {H_c}},
\end{equation}
which tend to Kasner indices $p_1$, $p_2$, $p_3$ when the cosmological evolution approaches the Kasner phase (see Fig. \ref{Fig2}, Fig. \ref{Fig3}, Fig. \ref{Fig4}). We can see from this plots that soon before reaching a cosmological singularity, these parameters tend to some constant values as it should be for the Kasner solution. Note that since we trace the evolution backward in time starting from $t=0$, the cosmological singularity appears 
at some time $t_0$ which depends on trajectory. The form of Kasner solution above have been written for the particular case $t_0=0$. Fig. \ref{Fig2} represent the time behaviour of $P_1$, the results for $P_2$ and $P_3$ are qualitatively the same. In Fig.~\ref{Fig3} and Fig.~\ref{Fig4} we plot time dependences of ~~$P_1+P_2+P_3$~~ and ~~${(P_1)}^2+{(P_2)}^2+{(P_3)}^2$~~ showing that they tend to their Kasner values for all studied trajectories. This justifies our statement that Kasner solution is an asymptotic solution for $T+f_0 T^N$ cosmology near a cosmological singularity.
     

\begin{figure}[hbtp]
~~~~~~~~~~~~\includegraphics[scale=0.68]{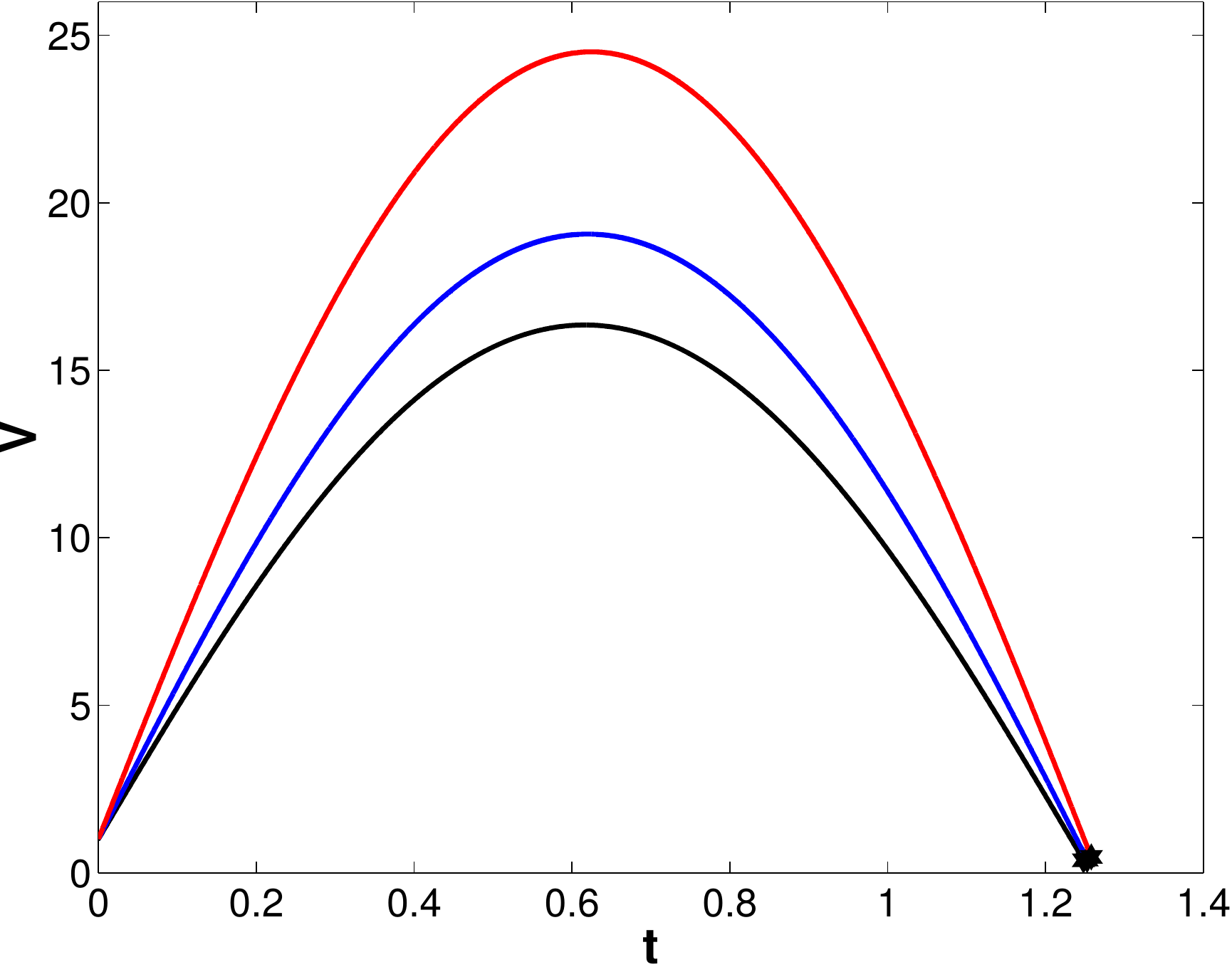}
\caption{The evolution of the volume factor $V(t)$ for $N=2$, $f_0=-\frac{1}{12}$. Initial values of $H_a(0)$, $H_b(0)$ are chosen as in Fig. \ref{Fig5} (see the corresponding colours of solid curves) and starting values of scale factors are $a(0)=1$, $b(0)=1$, $c(0)=1$ . The black star denotes the moment of a cosmological singularity in the future.}
\label{Fig5}
\end{figure}

\begin{figure}[hbtp]
\includegraphics[scale=0.44]{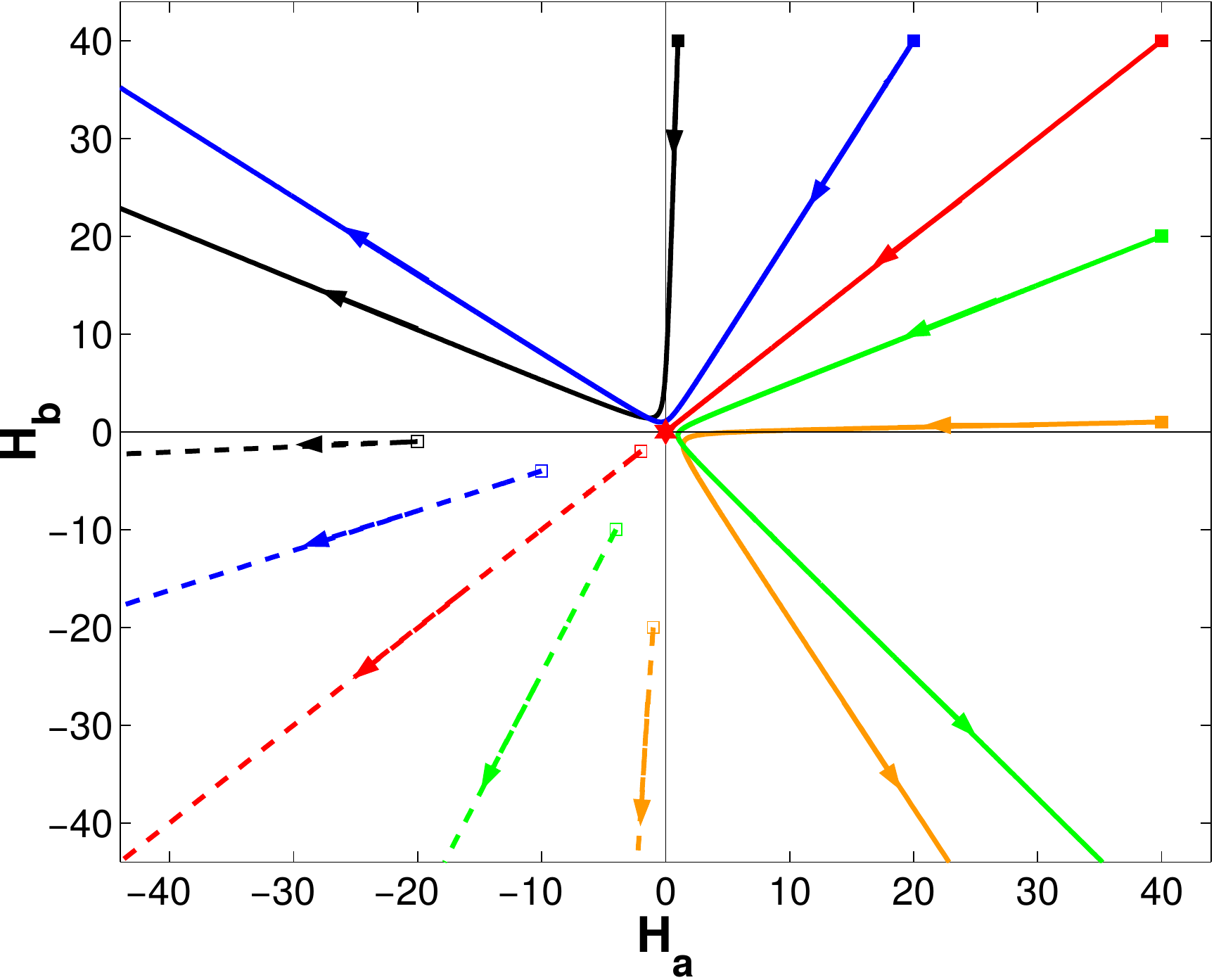}\qquad
\includegraphics[scale=0.44]{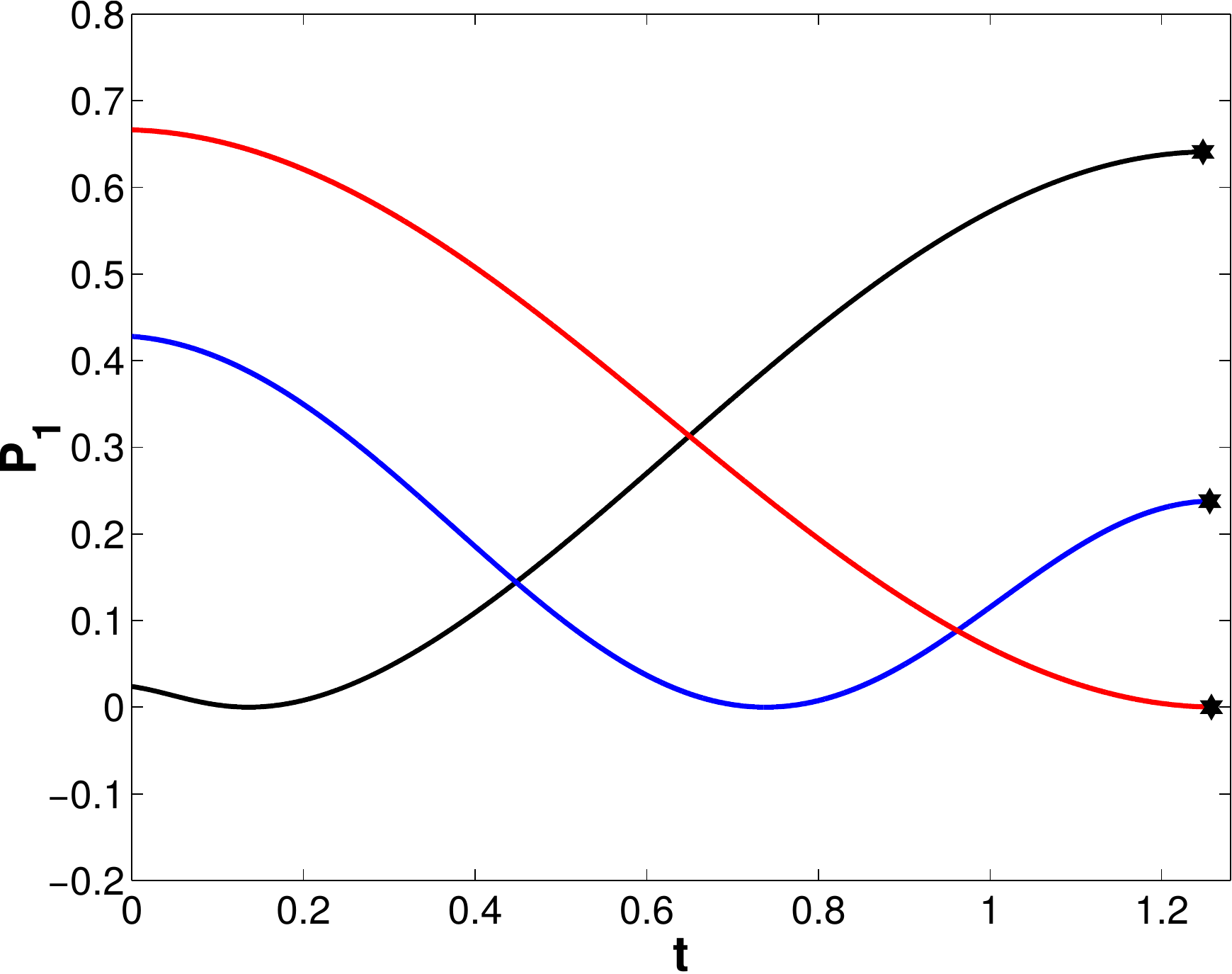} 
\caption{The phase portrait (left) and the evolution of the dimensionless parameter $P_1(t)$ (right) for $N=2$, $f_0=-\frac{1}{12}$. Initial values of $H_a(0)$, $H_b(0)$ are \textbf{1).-5).} for solid curves and \textbf{6).-10).} for dashed those:
\\\textbf{1).} $H_a(0)=1$, $H_b(0)=40$~~ (black),
~~\textbf{2).} $H_a(0)=20$, $H_b(0)=40$~~ (blue),
\\\textbf{3).} $H_a(0)=40$, $H_b(0)=40$~~ (red),
~~\textbf{4).} $H_a(0)=40$, $H_b(0)=20$~~ (green),
\\\textbf{5).} $H_a(0)=40$, $H_b(0)=1$~~ (orange),
~~\textbf{6).} $H_a(0)=-20$, $H_b(0)=-1$~~ (black),
\\\textbf{7).} $H_a(0)=-10$, $H_b(0)=4$~~ (blue),
~~\textbf{8).} $H_a(0)=-2$, $H_b(0)=-2$~~ (red),
\\\textbf{9).} $H_a(0)=-4$, $H_b(0)=10$~~ (green),
~~\textbf{10).} $H_a(0)=-1$, $H_b(0)=20$~~ (orange).
\\The red star denotes the point ~$(0; 0)$~ and the black star is the moment of a cosmological singularity in the future.}
\label{Fig6}
\end{figure}

\begin{figure}[hbtp]
\includegraphics[scale=0.44]{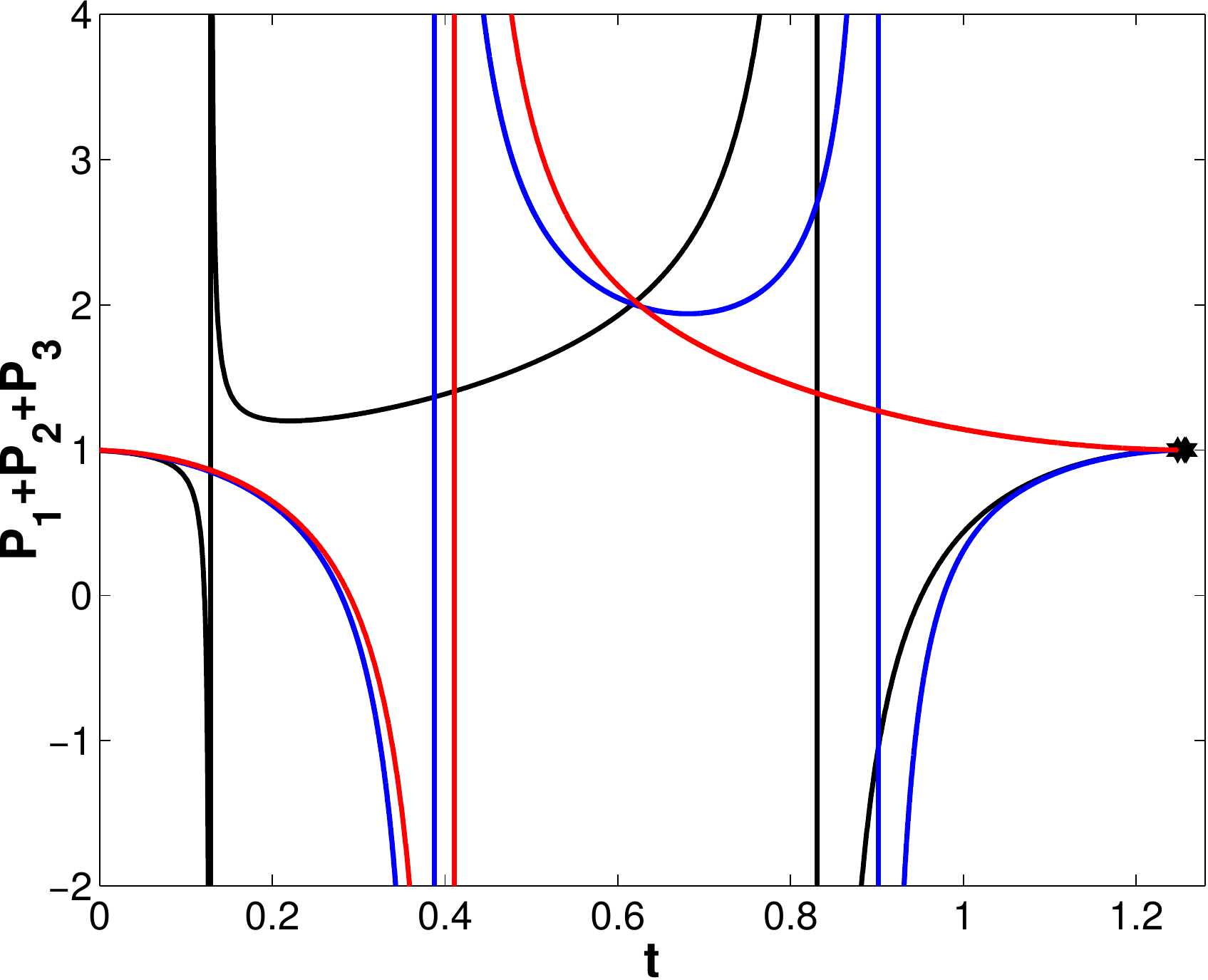}\qquad
\includegraphics[scale=0.44]{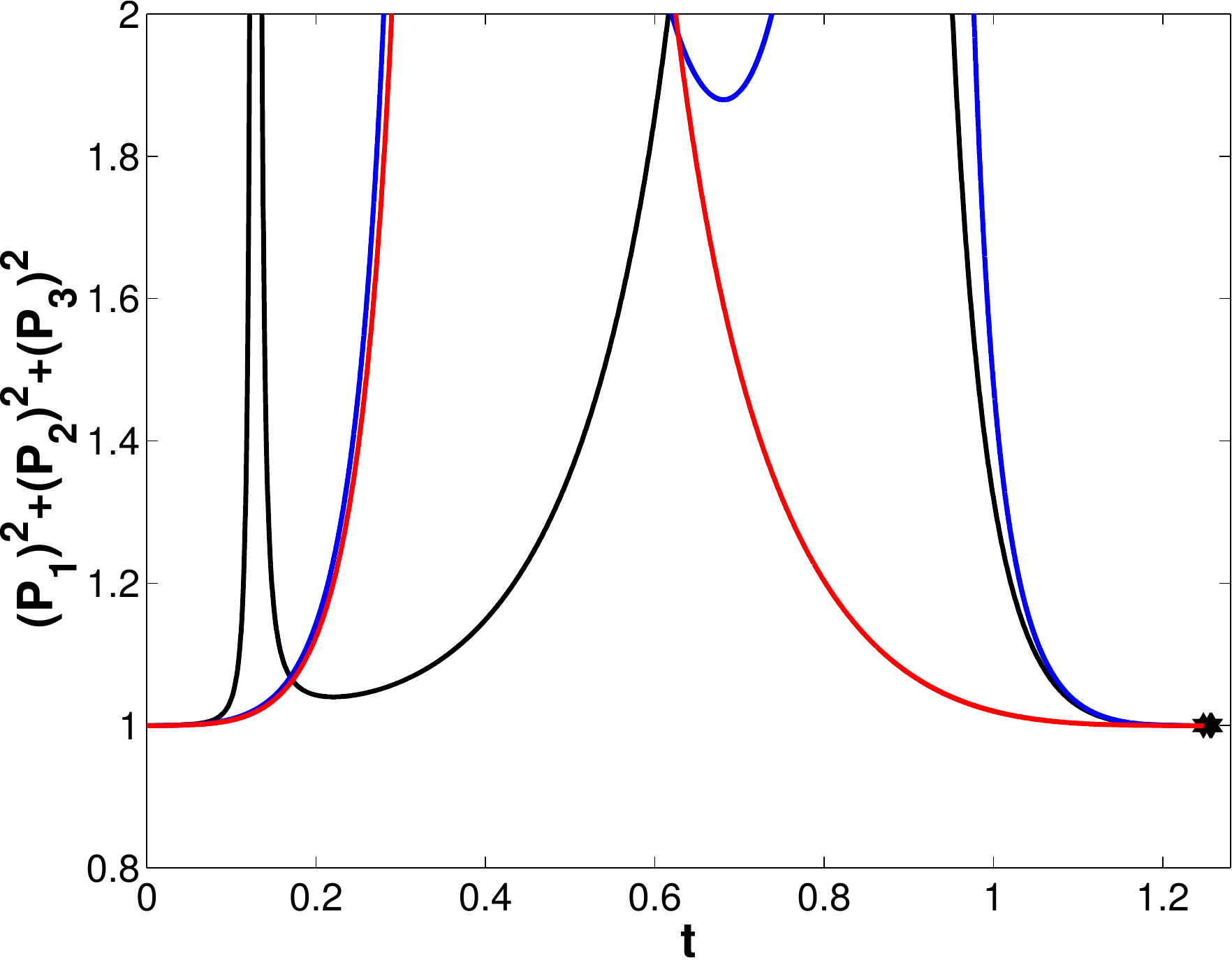} 
\caption{The evolution of sums $P_1+P_2+P_3$ (left) and ${(P_1)}^2+{(P_2)}^2+{(P_3)}^2$ (right) for $N=2$, $f_0=-\frac{1}{12}$. Initial values of $H_a(0)$, $H_b(0)$ are chosen as in Fig. \ref{Fig6} (see the corresponding colours of solid curves). The black star denotes the moment of a cosmological singularity in the future.}
\label{Fig7}
\end{figure}

Finally, we consider the case of $T=const>0$, where the de Sitter attractor does not exist. What happens here, can be seen from Fig. \ref{Fig5} where we plot the evolution of the volume of the Universe. In order to calculate the time dependences of the volume factor $V(t)\equiv a(t)b(t)c(t)$ we integrate the system (\ref{system1})-(\ref{system3}) with an addition three differential equations for scale factors: ~~$\dot a = H_a a$, ~~$\dot b= H_b b$, ~~$\dot c=H_c c$.~~ It appears that the volume reaches its maximum value at some time of the cosmological evolution, and after that the  Universe starts to recollapse. After that, the evolution ends in Big Crunch singularity, where the volume shrinks to zero at some time $t_0$. As near the Big Crunch singularity the absolute values of Hubble parameters grows infinitely, we can again neglect the constant $T$ in the equations of motion, which means that the metrics tends to the Kasner solution. In Fig. \ref{Fig6} and Fig. \ref{Fig7} we show numerically that for the branch $T=const>0$ the cosmological evolution begins and ends in Kasner solution as sums $P_1+P_2+P_3$ and ${(P_1)}^2+{(P_2)}^2+{(P_3)}^2$ tend to unit at late time. The saddle point in Fig. \ref{Fig6} (left) indicates the Minkowski vacuum solution $H_a=H_b=H_c=0$, any trajectory initially approaches this point, then, after reaching the maximal expansion point, departs from it. Initial Kasner indices differ from final those (see Fig. \ref{Fig6}. (right)). We have checked that while starting from arbitrary $H_a$, $H_b$ (and finding initial $H_c$ from the constraint equation) we get the same evolution from Kasner solution near Big Bang at some time $t_1$, then, after recollapse, Kasner solution near Big Crunch at some time $t_2$. In the intermediate region the meaning of parameters $P_1$, $P_2$ and $P_3$ is rather obscure, they can even diverges during cosmological evolution, since $\dot H_i$ entering denominator in the definition of corresponding $P_i$ may go through zero.


    To conclude we would like to underline in which points we agree and disagree with the paper \cite{Barrow2}. Since there are some subtleties, it is better to distinguish between the cases when the Kasner solution is an exact solution and an asymptotic solution. 

    We agree with \cite{Barrow2} that the Kasner solution can be an exact solution in $f(T)$ cosmology. However, it happens only on $T=0$ branch of vacuum solutions where the equations of motion coincide exactly with those for GR (any deviations from GR appears only when matter is taken into account). So that, the question of stability of Kasner solution in this situation is meaningless --- Kasner solution is the general vacuum solution on $T=0$ branch.

    On the contrary, Kasner solution is an asymptotic solution on the second branch. In this case it is possible to ask for its stability, and we agree with \cite{Barrow2} that it is unstable, and the cosmological evolution is generally directed from Kasner to de Sitter solution. However, in this situation Kasner solution can not be an exact solution of $f(T)$ cosmology.

So, the statement that the Kasner solution satisfies the equations of motion in $f(T)$ theory made in the Introduction of \cite{Barrow2}, and the statement that it is unstable (which is the main result of \cite{Barrow2}) are both correct only when applied to appropriate branches of solutions. They are not correct for other branch. However, in \cite{Barrow2} these statements have been formulated in general, without any reference to particular branches.

Moreover, our study does not support the claim of \cite{Barrow2} that Kasner solution is a saddle. On the contrary, we can see from Figs. \ref{Fig1}-\ref{Fig4} that on the $T \neq 0$ branch, where the Kasner solution is unstable, it represents a source in an expanding Universe. This means that it is stable into the past, as well as it is an attractor for a contracting Universe.

Our results can be understood in another way if we note  that vacuum equations of motion of the models $f(T)=T+f_0T^N$ for the branch $T=const\neq0$ coincide with those of anisotropic Bianchi I models of GR with cosmological constant $\Lambda$, where $T>0$ corresponds to $\Lambda<0$ ($T<0$ corresponds to $\Lambda>0$). Therefore, a cosmological evolution is directed from Kasner to de Sitter solution in $T<0$ case, and from one Kasner solution to the point of maximal expansion (see Fig. \ref{Fig7}) and back to another Kasner solution in the case of positive $T$. 

    As for the statement of \cite{Rodrigues} that a vacuum solution can exist only for $f(T)=\sqrt{-T}$ theory, to our mind, it follows from an incorrect interpretation of the constraint equation (\ref{constraint}). This equation may be considered as a {\it differential} equation which should get us such $f(T)$ that the constraint is valid for all $T$. In this interpretation authors of \cite{Rodrigues} are correct. However, this equation can be considered as an {\it algebraic} one which give us some {\it particular} $T$ for which the constraint equation has solutions. This situation is known in $f(R)$ cosmology. The equation for vacuum de Sitter solution $2f(R)-Rf'(R)=0$ when considered as a differential equation gives particular $f(R)=R^2$, for which de Sitter solution with any $R$ exists. On the contrary, the same equation being algebraic equation gives for other $f(R)$ some particular $R$ for which de Sitter solution exists. In $f(R)$ theory this is restricted by de Sitter solutions only, because in general situation there are other derivative terms. In $f(T)$ the situation is general. If we search for any vacuum solution, there is no need for this solution to be valid for any $T$, so the constraint equation should be considered as an algebraic one.

    We remind a reader that all results of the present paper have been obtained under suggestion that the tetrad $(1,a,b,c)$ is the proper tetrad for Bianchi I cosmology. Our results showing that the Kasner solution is either the exact general solution or asymptotic solution stable to the past indicate that it is reasonable to search for an analogue of BKL oscillations in anisotropic $f(T)$ cosmology with spatial curvature. However, this needs to identify proper tetrads for other Bianchi metrics (or, in another formulation of the theory, correct non-zero spin connections for these cases), which is a separate interesting and currently unsolved problem of $f(T)$ gravity.

\section*{Acknowledgements}
~~~~The work was supported by RSF Grant \textnumero 16-12-10401 and
by the Russian Government Program of Competitive Growth of Kazan Federal University.

\end{document}